\documentclass[twocolumn,noshowpacs,pre,preprintnumbers,superscriptaddress,amsmath,amssymb,floatfix]{revtex4}
\usepackage[caption=false]{subfig}
\usepackage{graphicx}
\usepackage{color}
\usepackage{placeins}
\usepackage[normalem]{ulem}
\usepackage[colorlinks=true,citecolor=blue,linkcolor=blue,urlcolor=black]{hyperref}

\newcommand{\vbax}{\lvert\vb{x}\rvert}
\newcommand{\prw}{\mathcal{P}_{ij}^{\mathrm{SE}}}

\usepackage{physics}

\begin{document}

\title{Percolation in heterogeneous spatial networks with long-range interactions}
	\author{Guy Amit}
    \email[]{guy.amit@biu.ac.il}
	\affiliation{Department of Physics, Bar-Ilan University, Ramat Gan, Israel}
    \author{Dana Ben Porath}
    \thanks{Formerly Dana Vaknin}
	\affiliation{Department of Physics, Bar-Ilan University, Ramat Gan, Israel}
	\author{Sergey V. Buldyrev}
	\affiliation{Department of Physics, Yeshiva University, New York, USA}
    \author{Amir Bashan}
    \email[]{amir.bashan@biu.ac.il}
	\affiliation{Department of Physics, Bar-Ilan University, Ramat Gan, Israel}
	\date{\today}









\begin{abstract}
We study the emergence of a giant component in a spatial network where the distribution of the metric distances between the nodes is scale-invariant, and the interaction between the nodes has a long-range power-law behavior.
The nodes are positioned in the metric space using a L\`evy flight procedure, with an associated scale-invariant step probability density function, and is then followed by a process of connecting each pair of nodes with a probability function that depends on the distance between them. 
A natural way to analyze the system is to consider the total probability for an edge between steps in term of their indexes, by summing over their possible positions. By doing so, a correspondence is found between this model and a model of percolation in a one-dimensional lattice with long-range interactions, which allows the identification of the conditions for which a percolation transition is possible. We find that the emergence of a giant component and percolation transitions is determined by a complicated phase diagram, that exhibits a transition from weak long-range interactions to strong long-range interactions. 
\end{abstract}


\maketitle

\textit{Introduction.}--In this work we ask and answer the following question: If the nodes of a spatial network are embedded in a metric space in a scale-invariant manner, 
and the interactions between the nodes are governed by a power-law relationship, under what conditions will a giant component emerge?

Spatial networks \cite{barthelemy2011spatial} are a class of networks where the nodes are embedded in a metric space, meaning that the nodes have associated coordinates that allow to define distances between them. Typically, the edges between the nodes are constructed such that a pair of nodes with a small distance between them have a larger probability of being connected with an edge compared with nodes which are far apart. Many real-world systems can be modeled as spatial networks, including transportation systems \cite{zeng2019switch, zeng2020multiple, cogoni2021stability} power-grids \cite{yang2017small}, communications \cite{lambiotte2008geographical}, urban streets \cite{wu2004urban, zhang2019scale} infrastructure \cite{rinaldi2001identifying, chang2009infrastructure, hines2010topological}  and brain activity \cite{fornito2016fundamentals, breakspear2017dynamic}.
It is of particular importance to know whether a network contains a Giant Component (GC), i.e., a connected cluster of nodes whose size, $M$, remains greater than $cN$, where $0<c\leq1$ is some constant, for any $N\to\infty$. The existence of a GC in a network signifies the ability of the network to function properly \cite{molloy1998size, strogatz2001exploring, newman2001random, newman2002spread}.  

Here, we consider \textit{scale-invariant} spatial networks in term of the spatiality, i.e., the metric distance between the nodes follows a power-law distribution. To mitigate confusion, we emphasize that when we say scale-invariant we refer to the metric distance between the nodes, and we reserve the term scale-free to the degree distribution \cite{aldous2013true}. 
We model the spatial placement of the nodes in the network using a L\`evy flight process, a generalization of Brownian motion, that is characterized by infinite variance, which was used to model many physical phenomena \cite{shlesinger1995levy}, and in particular, foraging problems \cite{viswanathan2000levy, boyer2006scale, viswanathan2008levy, codling2008random, sims2008scaling, james2011assessing}. 
%


In addition to the scale-invariant placement of the nodes, the interactions are Long-Ranged (LR), i.e., the bond probability is taken from a slowly decaying power-law function of the distance between them. 
Examples of long-range interactions in physical systems can be found in a wide range of subjects, such as astrophysics, plasma physics, hydrodynamics, atomic physics, and nuclear physics \cite{campa2009statistical, campa2014physics, defenu2021long}. Long-range interactions in complex networks were also used to model disease spreading \cite{solomon2000social,newman2002spread}, social aggregation \cite{solomon2000social, frasco2014spatially} and financial transactions \cite{stauffer2001percolation}. 

The percolation problem of geometric networks with Short-Range (SR) interactions has been the subject of many recent studies \cite{gilbert1961random, dall2002random, penrose2003random, xie2016scale, dettmann2016random, murphy2018geometric, plaszczynski2022levy, grosspercolation}. In addition, the seminal problem of percolation on a lattice with long-range interactions has also been extensively studied \cite{kac1969critical, dyson1969existence, stephen1981percolation, schulman1983long, zhang1983long, imbrie1988intermediate, aizenman1987sharpness, newman1986one, coppersmith2002diameter, moukarzel2006percolation, d2016phase, gori2017one}. 
However, real-world spatial networks are typically heterogeneous (in marked contrast with lattices) and can have long-range interactions. Therefore, it is important to understand the emergence of a GC in scenarios that involve both of these conditions.

\begin{figure}%
\centering
\includegraphics[width=0.5\textwidth]{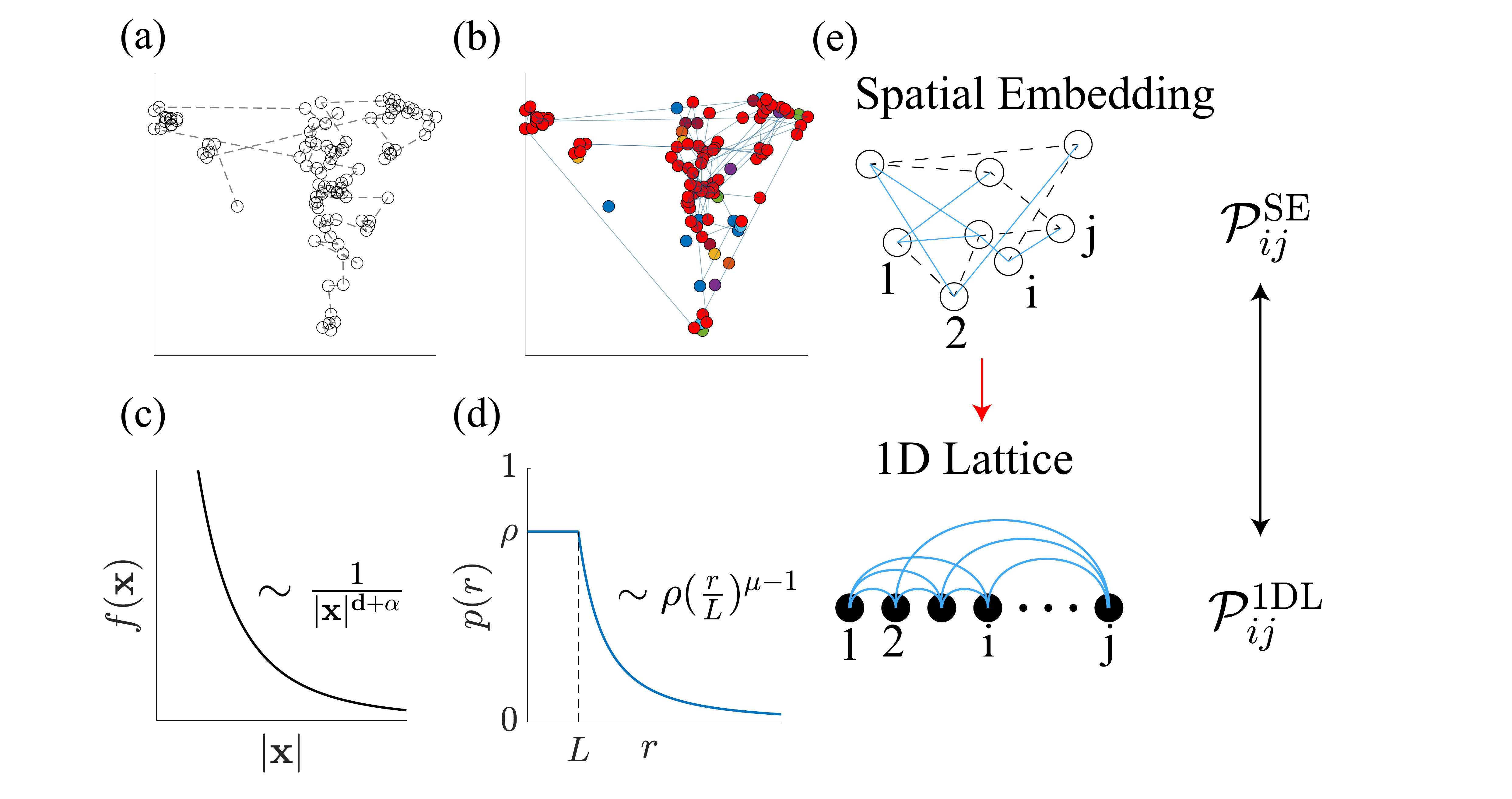}
\caption{\textbf{Graphical description of the model}. (a) A L\`evy flight process positions the nodes in the $d$-dimensional metric space. The step-lengths (black dashed line) are taken from a scale-invariant probability density function $f(\vb{x})$, with an associated stability parameter $\alpha$. (b) The nodes are then connected with edges according to a probability function $p(r)$ which depends on the distance between them and has an associated long-range interaction parameter $\mu$. The different colored nodes represent different separated clusters. The red cluster is the largest connected component (c) Visualization of the function $f(\vb{x})$ with its long-tail behavior that depends on the dimensionality $d$ and on the stability parameter $\alpha$. (d) Visualization of the function $p(r)$ with its associated length $L$ and the long-range interaction parameter $\mu$. (e) We calculate the total probability that step $i$ is connected to step $j$ by summing over all possible locations of step $j$. We find a correspondence between the existence of a GC in this problem and the problem of percolation in a one-dimensional lattice with long-range interactions.}\label{fig1_model}
\end{figure}

\textit{Model.}--We begin by detailing the process of building the network. A graphical description is presented in Fig. \ref{fig1_model}.
The $N$ nodes of the network are embedded in a $d$-dimensional space with a radially-symmetric L\`evy flight process, such that in each step the direction is picked from a uniform distribution, and the step length $r$ is chosen from a distribution that, for large $r$, has a power-law behavior $\sim\frac{1}{r^{1+\alpha}}$. The probability density function of the position of the steps is therefore given by \cite{ross1996stochastic}
\begin{equation}\label{fx}
    f(\vb{x})\sim\frac{1}{\vbax^{d + \alpha}}, \quad \frac{\vbax}{\gamma}\gg 1,
\end{equation}
where $\vb{x}\in\mathbb{R}^d$, $\alpha>0$ is the stability parameter, and the probability density function has an associated scaling factor $\gamma$ such that $\vbax/\gamma$ is dimensionless. When $\alpha<2$ the distances between the nodes is scale-invariant and a mean distance is undefined.

\hfill\newline
Then, each pair of nodes, $i$ and $j$, are connected with an edge with a probability that depends on the distance between them $p(\lvert\vb{x}_i-\vb{x}_j\rvert)$. We choose the following long-range probability function,
\begin{equation}\label{pmu}
    p(r_{ij}) = 
    \begin{cases}
    \rho, & r_{ij}\leq L \\
    \rho\left(\frac{r_{ij}}{L}\right)^{\mu-1}, & r_{ij}>L
    \end{cases}, \quad\mu\leq 1
\end{equation}
where $r_{ij} = \lvert\vb{x}_i - \vb{x}_j\rvert$, $L>0$ and the parameter $0<\rho\leq1$ is the probability that two nodes, which are at a distance $r_{ij}\leq L$ apart, are connected with an edge. The parameter $\mu$ is responsible for long-range interactions (see Fig. \ref{fig1_model}(d)). When $\mu=1$ the probability for an edge between any two nodes is equal to $\rho$ independently of the distance between them, i.e., an Erd\H os–R\`enyi network. 
When $\mu=-\infty$ the function $p(r_{ij})$ is given by
\begin{equation}\label{pgeo}
    p(r_{ij}) =   
    \begin{cases}
    \rho, & r_{ij}\leq L \\
    0, & r_{ij}>L
    \end{cases}, \quad \mu=-\infty.
\end{equation}
In this case, the nodes are connected with a probability $\rho$ if the distance between them is less than or equal to $L$, and 0 otherwise, i.e., a soft geometric network \cite{gilbert1961random, dettmann2016random, penrose2016connectivity, wilsher2020connectivity, erba2020random, plaszczynski2022levy}. 

\textit{Results.}--The probability for an edge between node $i$ and node $j$ can be calculated explicitly. We denote the total probability that step $i$ is connected with an edge to step $j$ after $\lvert i - j\rvert$ steps with $\mathcal{P}_{ij}^{\text{SE}}$ (where SE stands for Spatial Embedding).
To find $\mathcal{P}_{ij}^{\text{SE}}$, we sum over all possible positions of node $j$, $\vb{x}_j$, whilst treating the position of node $i$, $\vb{x}_i$, as a constant (assuming $j>i$ without loss of generality), i.e.,
\begin{widetext}
\begin{align} \label{pconv}
    \mathcal{P}_{ij}^{\text{SE}}(\vb{x}_i) = \int\dots\int p(\lvert\vb{x}_i - \vb{x}_j\rvert) f(\vb{x}_j-\vb{x}_{j-1})\dots f(\vb{x}_{i+1}-\vb{x}_{i})\dd[d]{\vb{x}_j}\dots\dd[d]{\vb{x}_{i+1}}
    =(p\,{\textstyle*d}\,f^{{*d}\lvert i-j\rvert})(\vb{x}_i),
\end{align}
\end{widetext}
where ${*d}$ is the $d$-dimensional convolution, $f^{{*d}\lvert i-j\rvert}$ is a $\lvert i-j \rvert$-times $d$-dimensional convolution power of $f(\vb{x})$, and the integrals are over the entire volume of the $d$-dimensional space. 


To evaluate (\ref{pconv}), we apply a Fourier transform \cite{ross1996stochastic}. We define $\tilde{f}(\boldsymbol{\nu})$ and $\tilde{p}(\boldsymbol{\nu})$ as the Fourier transform of $f(\vb{x})$ and $p(\lvert\vb{x}\rvert)$, respectively.
\begin{align}
\begin{split}
    &\tilde{f}(\boldsymbol{\nu}) = \int_{\mathbb{R}^d} f(\vb{x}) e^{-i\boldsymbol{\nu}\cdot\vb{x}}\dd[d]{\vb{x}},\\
    &\tilde{p}(\boldsymbol{\nu}) = \int_{\mathbb{R}^d} p(\vb{x}) e^{-i\boldsymbol{\nu}\cdot\vb{x}}\dd[d]{\vb{x}},
    \end{split}
\end{align}
where $\boldsymbol{\nu}\in\mathbb{R}^d$.
With these definitions at hand, the Fourier transform of $\mathcal{P}_{ij}^{\text{SE}}(\vb{x}_i)$ is
\begin{equation}
    \tilde{\mathcal{P}}_{ij}^{\text{SE}}(\boldsymbol{\nu}) = \tilde{p}(\boldsymbol{\nu})(\tilde{f}(\boldsymbol{\nu}))^{\lvert i-j\rvert}.
\end{equation}
Since $f(\vb{x})$ and $p(\lvert\vb{x}\rvert)$ are independent of the step index, when calculating the probability $\mathcal{P}_{ij}^{\text{SE}}$ we can set $\vb{x}_i=0$ at the origin without loss of generality, i.e., $\mathcal{P}_{ij}^{\text{SE}}=\mathcal{P}_{ij}^{\text{SE}}(\vb{x}_i = 0)$.
Finally, $\mathcal{P}_{ij}^{\text{SE}}$ can be calculated via the inverse Fourier transform
\begin{equation}\label{inversefd}
    \mathcal{P}_{ij}^{\text{SE}} = \frac{1}{(2\pi)^d}\int_{\mathbb{R}^d} \tilde{p}(\boldsymbol{\nu})(\tilde{f}(\boldsymbol{\nu}))^{\lvert i-j\rvert}\dd[d]{\boldsymbol{\nu}}.
\end{equation}
Using (\ref{inversefd}) we will find the conditions for which percolation is possible. Before continuing, we will do a short interlude and talk about percolation on a one-dimensional lattice with long-range interactions.

Consider a one-dimensional lattice, where nodes are embedded on the lattice points, and each pair of nodes $i$ and $j$ are connected with an edge with a probability $\mathcal{P}_{ij}^\text{1DL}$ (see Fig. \ref{fig1_model}(e); 1DL stands for 1-Dimensional Lattice). A well studied scenario of this problem involves long-range interactions of the form 
\begin{equation}\label{PLP}
    \mathcal{P}_{ij}^\text{1DL} = \frac{C}{\lvert i-j \rvert^{1+\sigma}}, \quad 0<C\leq 1.
\end{equation}
With this form of interactions, GC exists under the following conditions \cite{gori2017one}. For $\sigma\geq1$, a GC exists if and only if $C=1$. This means that a GC exists only if nearest-neighboring nodes are guaranteed to be connected with an edge. For $0<\sigma< 1$, a GC exists if $C$ is larger than some critical value $C_c$. Schulman \cite{schulman1983long} showed that \mbox{$C_c\geq1/2\zeta(1+\sigma)$} where $\zeta(x)$ is the Riemann zeta function, $\zeta(x)=\sum_{n=1}^\infty \frac{1}{n^x}$, while Gori et al \cite{gori2017one} found precise numerical estimates of $C_c$. For $\sigma\leq0$ a GC exists for arbitrarily small $C$.

Coming back to our random walk model, although the nodes are embedded in a $d$-dimensional space, Eq. (\ref{inversefd}) defines the probability that two nodes are connected in term of their indexes, not their position. If the long-term behavior of (\ref{inversefd}) is similar to (\ref{PLP}), we can find the conditions under-which a GC exists, by corresponding the random walk problem in $d$ dimensions to the percolation threshold of a one-dimensional lattice. Indeed, with the use of asymptotic expansion, we find that the SE model has a similar power-law probability of $\mathcal{P}_{ij}^\text{1DL}$, with $\sigma$ being a function $\alpha$, $\mu$ and $d$, and $C$ is proportional to $\rho$,
\begin{equation}
    \mathcal{P}_{ij}^{\text{SE}} \Longleftrightarrow \mathcal{P}_{ij}^\text{1DL}, \quad \text{for }\lvert i-j\rvert\gg 1.
\end{equation}
In other words, although the nodes are embedded in a $d$-dimensional space, the probability for an edge between any two nodes depends simply on the order in which the nodes appeared in the L\`evy flight path, which can be thought of as a one dimensional lattice such that each lattice point represents an index of the path. 
In this correspondence, the factor $\rho$ from the random walk model (\ref{inversefd}) takes the role of the constant $C$ from the 1D lattice model (\ref{PLP}). We expect that for certain choices of $\mu$, $\alpha$ and $d$, a critical value, $0<\rho_c<1$, can exist such that a GC will emerge as $\rho$ is increased from $\rho<\rho_c$ to $\rho>\rho_c$.

We solve Eq. (\ref{inversefd}) with asymptotic expansion, i.e., for large $\lvert i-j\rvert\gg 1$. The full derivation is presented in the Supplementary Material (SM). A key step in the analysis is that, due the Generalized Central Limit Theorem \cite{gnedenko1954limit}, the term $(\tilde{f}(\boldsymbol{\nu}))^{\lvert i-j\rvert}$ on the r.h.s. of (\ref{inversefd}), for $\lvert i-j\rvert\gg 1$, is given by
\begin{small}
\begin{align} \label{CLT}
\begin{split}
    (\tilde{f}(\boldsymbol{\nu}))^{\lvert i-j\rvert}&=\exp\left\{-\lvert\gamma\nu\rvert^\alpha\lvert i-j\rvert\right\},\quad \text{for } 0<\alpha<2, \\
    (\tilde{f}(\boldsymbol{\nu}))^{\lvert i-j\rvert}&=\exp\left\{-\lvert\gamma\nu\rvert^2\lvert i-j\rvert\right\},\quad \text{for } \alpha\geq2,
\end{split}
\end{align}
\end{small}
where $\nu=\lvert\boldsymbol{\nu}\rvert$.

For the geometric case \mbox{($\mu=-\infty$)} the first order term of $\prw$ is given by
\begin{equation} \label{geodd}
    \mathcal{P}_{ij}^{\text{SE}}(\mu=-\infty)\sim\frac{d}{2^{d}\alpha}\frac{\left(\frac{L}{\gamma}\right)^d\Gamma\left(\frac{d}{\alpha}\right)}{(\Gamma\left(\frac{d}{2}+1\right))^2}\times\frac{\rho}{\lvert i-j\rvert^{d/\alpha}}.
\end{equation}
This relation already hints to the connection with the 1D lattice case, as we can observe that the factor \mbox{$1 + \sigma$} in \mbox{(\ref{PLP})} is replaced here with $d/\alpha$. Indeed, even in the geometric case, which is characterized by short-range interactions in the $\mathbb{R}^d$ metric space, there are long-range interactions in the index-space. However, for this case we found no numerical evidence of percolation transition or a critical value $0<\rho_c<1$. In general $\rho_c>1$, unless $\alpha\geq d$, in-which-case $\rho_c=0$ (which coincides with a return probability of 1 \cite{ross1996stochastic}). For further discussion see the SM.

In the long-range case ($\mu>-\infty$), we find the following form of $\prw$
\begin{equation} \label{mudd}
\begin{cases}
\begin{alignedat}{2}
    &\mathcal{P}_{ij}^\text{SE}\sim \frac{\rho}{\lvert i-j\rvert^{\frac{1-\mu}{\alpha}}},\quad &&\text{for } 1-d<\mu<1, \\
    &\mathcal{P}_{ij}^\text{SE}\sim \frac{\rho}{\lvert i-j\rvert^{d/\alpha}},\quad &&\text{for }\mu<1-d, 
\end{alignedat}
\end{cases}
\end{equation}
where $\alpha$ is replaced by $2$ if $\alpha>2$ due to (\ref{CLT}). The derivations are presented in the SM.

\begin{figure}[h]%
\centering
\includegraphics[width=0.45\textwidth, trim={1cm 1cm 0 0}]{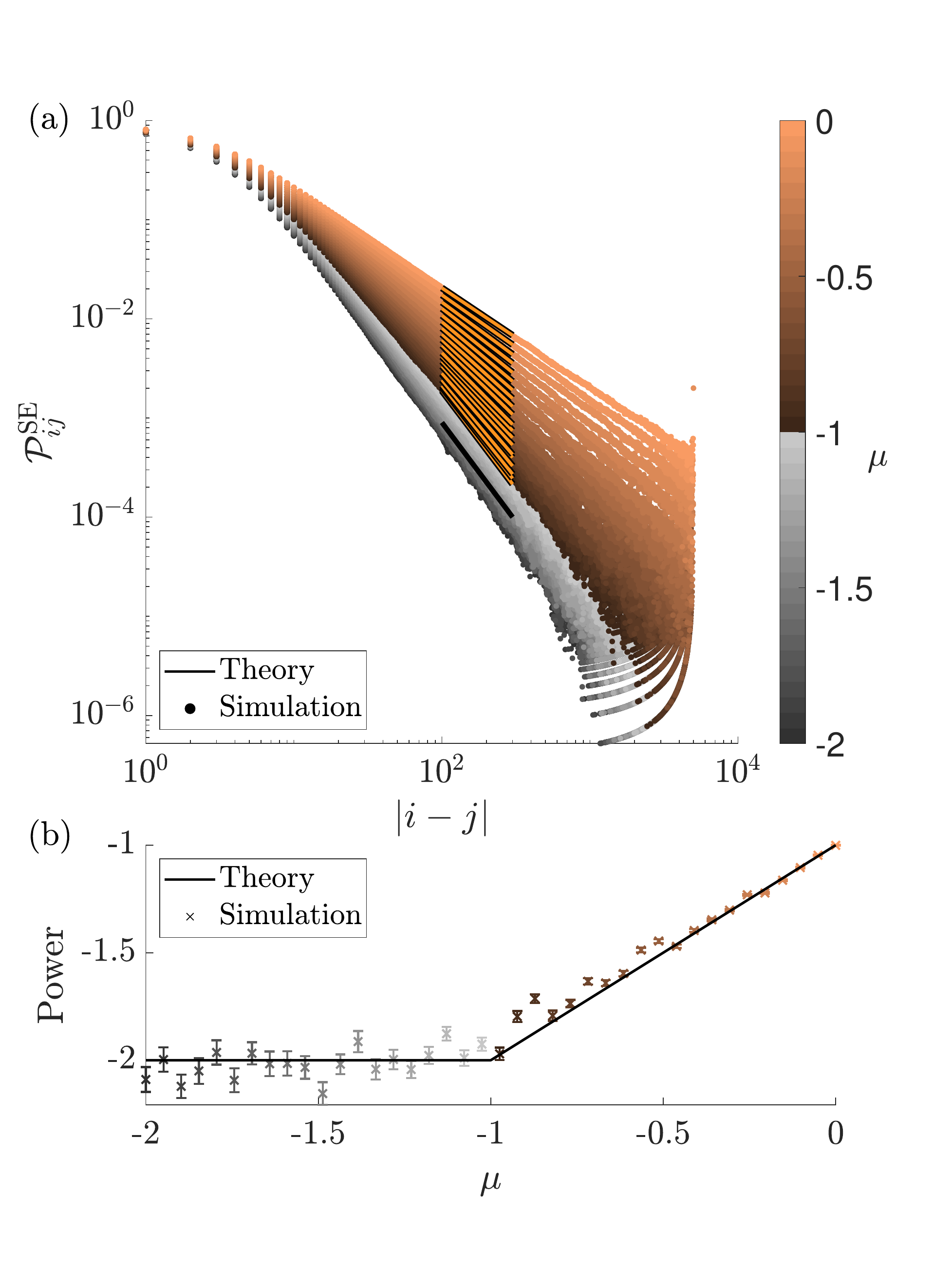}
\caption{\textbf{Transition from weak LR to strong LR interactions}. (a) The probability for an edge, $\prw$, as a function of $\lvert i-j\rvert$ for different values of $\mu$. Here $\alpha = 1$ and $d=2$. In this case, from Eq. (\ref{mustar}), $\mu^*=-1$. For $\mu>\mu^*$ the power behavior of the probability for an edge depends on $\mu$ and is equal to $(1-\mu)/\alpha$ (the strong LR case). For $\mu<\mu^*$ the power behavior is independent of $\mu$ and is equal to $d/\alpha$ (the weak LR case). Each value of $\mu$ is averaged over 500 realizations. (b) Comparison between the numerically calculated power of $\prw$ and the theory. For each line in (a) we calculated a simple linear fit. The bar-lines show the confidence intervals. Here $N=10^4$.}\label{figure6}
\end{figure}

\begin{figure}[h]%
\centering
\includegraphics[width=0.5\textwidth, trim={0cm 3cm 0 0}]{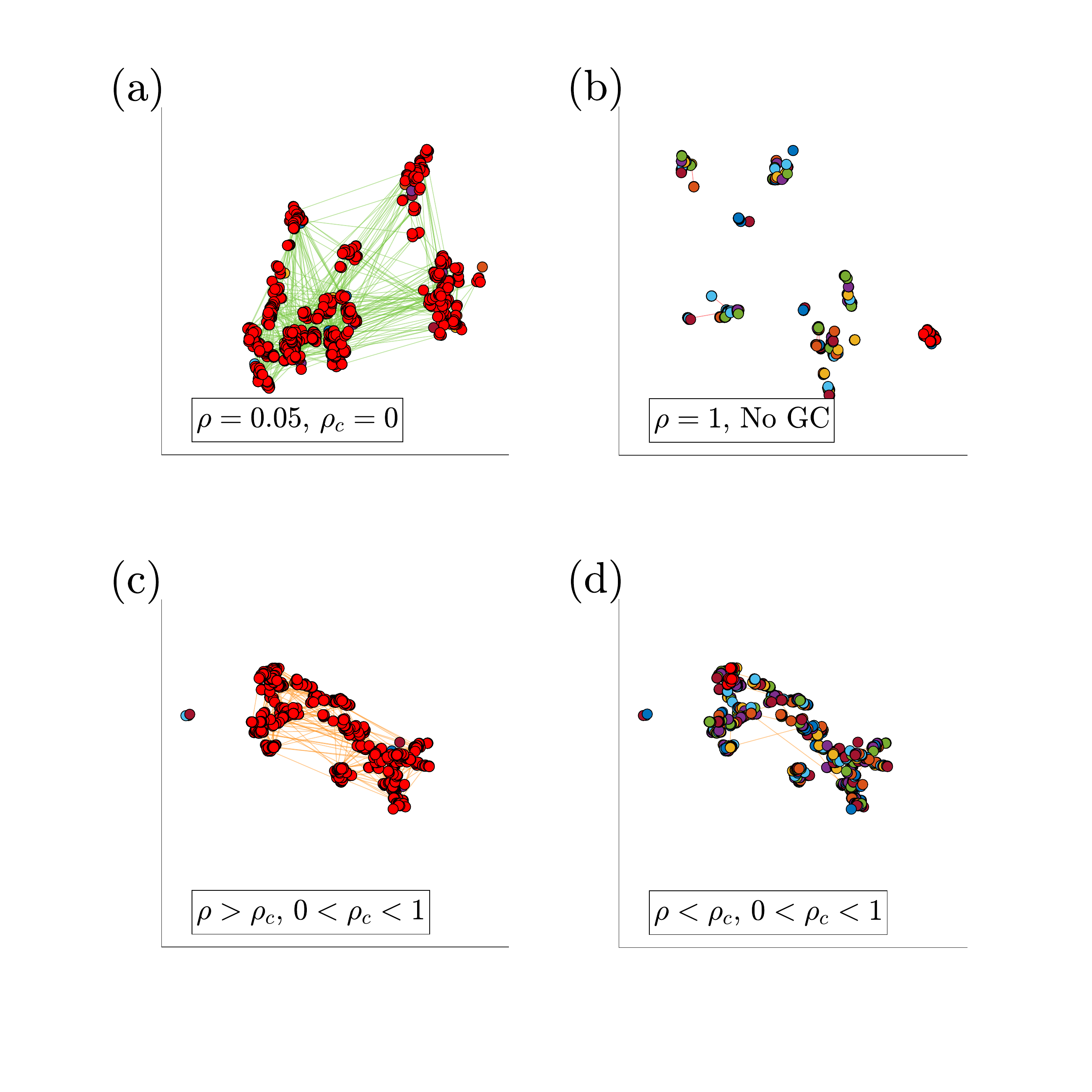}
\caption{\textbf{Examples of networks with and without phase transitions}. (a) A case of $\alpha=1.5$ and $\mu=-0.25$. These parameters correspond to the mean field universal behavior region (green area in Fig. \ref{figure9}), which has $\rho_c=0$ (i.e., no phase transitions). Here we chose $\rho=0.05$. (b) A case of $\alpha=0.8$ and $\mu=-0.8$. In this region a GC is guaranteed not to exist ($\rho_c>1$, red area in Fig. \ref{figure9}). Here we chose $\rho=1$. (c) and (d) Cases of $\alpha=1.2$ and $\mu=-0.8$. This is a region where phase transitions are possible, and in the strong LR case (grey area in Fig. \ref{figure9}). For (c) $\rho=0.5$ and for (d) $\rho=0.05$, the system exhibits a phase transition (i.e., $0.05<\rho_c<0.5$). In all cases, the bright red cluster is the largest component.}\label{figure7}
\end{figure}

\begin{figure}[h]%
\centering
\includegraphics[width=0.5\textwidth, trim={1cm 1cm 0 0}]{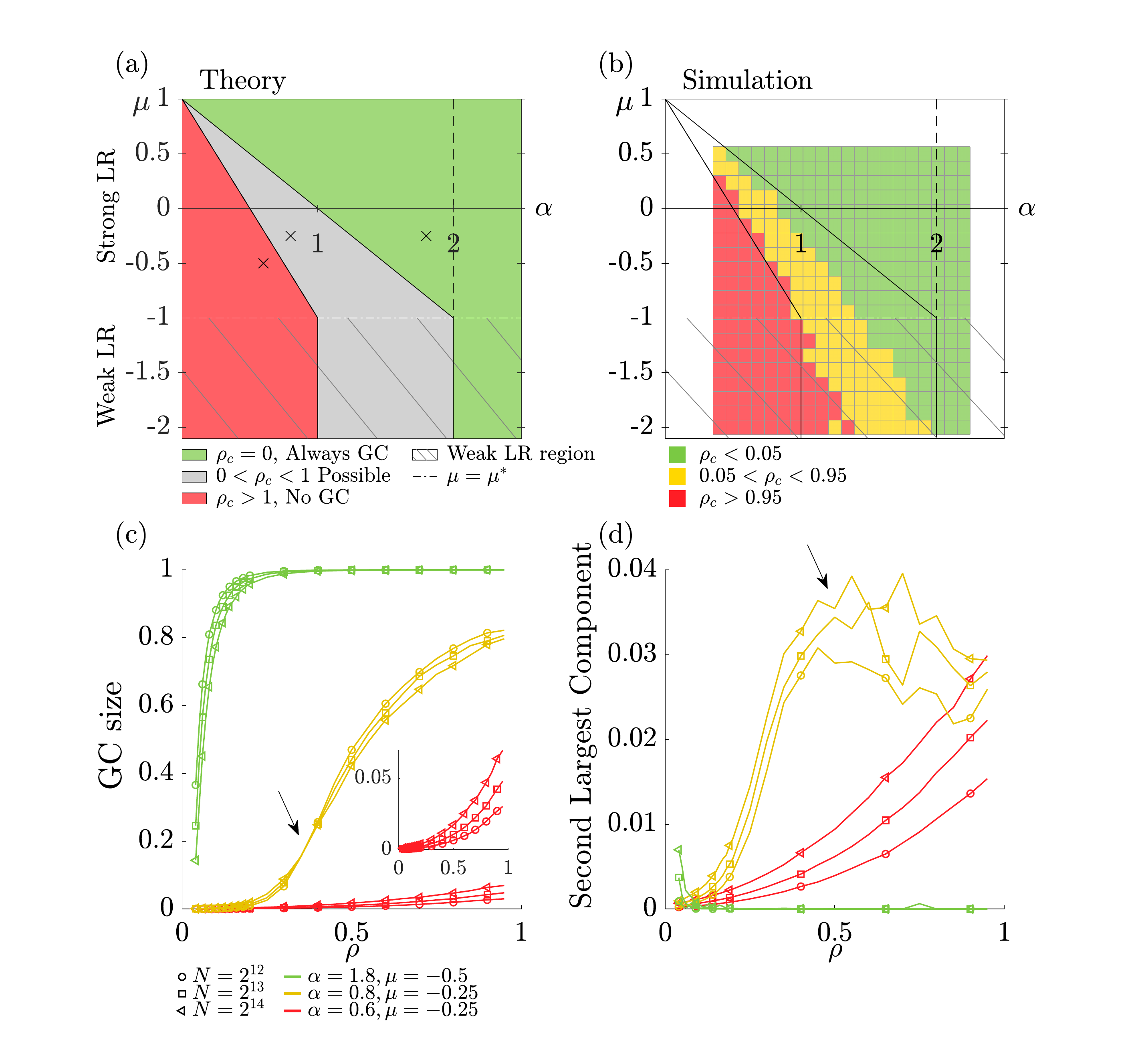}
\caption{\textbf{Phase diagram for the case of $d=2$}. (a) Theoretical prediction of the different regions. The diagram shows the different regions where a GC can and cannot exist (green and red respectively), and where percolation transition is possible (grey), as well as the strong and weak LR regions (above and below the line $\mu=\mu^*$, respectively). (b) Simulated phase diagram, colored according to detection of phase transitions with the use of the maximal values of the second largest component. Squares are colored green if $\rho_c<0.05$, red if $\rho_c>0.95$ and gold if $0.05<\rho_c<0.95$. Here $N=100$ and the number of realizations is 100. (c) Simulations of the size of the GC for three cases of $\mu$ and $\alpha$ marked by x in the phase diagram plot, for various number of nodes $N$, averaged over 1000 realizations (red: $\alpha=0.6$, $\mu=-0.25$; orange: $\alpha=0.8$, $\mu=-0.25$; green: $\alpha=1.8$, $\mu=-0.5$). The black arrow shows the crossing of the size of the which is near the critical value $\rho_c$ (inset: zoom in on the red curves). (d) Simulations of the size of the second largest component. The arrow marks the maximal values, which is near the critical value $\rho_c$.}\label{figure9}
\end{figure}

Let us consider what Eq. (\ref{mudd}) means. 
The system behaves qualitatively differently as $\mu$ is changed from \mbox{$\mu>\mu^*$} to \mbox{$\mu<\mu^*$}, where
\begin{equation}\label{mustar}
    \mu^* = 1-d.
\end{equation}
When $\mu>\mu^*$ the dominating power of the long-range interaction is of the form $\frac{1-\mu}{\alpha}$. When $\mu<\mu^*$ the dominating power is of the form $d/\alpha$, exactly like in the geometric, short-range, case (\ref{geodd}). The case of $\mu>\mu^*$ is typically called \textit{strong long-range interactions}, and of $\mu<\mu^*$ \textit{weak long-range interactions} \cite{fisher1972critical, defenu2020criticality}. The transition from strong LR to weak LR in the case of $d=2$ is presented in Fig. \ref{figure6}. Corresponding examples of networks, which exhibit cases where phase transitions are possible and impossible, are presented in Fig. \ref{figure7}. 
A full picture of the phase diagram in the $d=2$ is presented in Fig. \ref{figure9}, and for other values of $d$ in the SM.

First, let us consider the weak LR region ($\mu<\mu^*$). In Fig. \ref{figure9} it is the region below the dotted line of $\mu=-1$. In this case, we compare our model with the 1D lattice model (\ref{PLP}), by matching the powers of $\lvert i-j\rvert$ and find that
\begin{equation}\label{weaklr}
    1+\sigma = d/\alpha, \quad \text{for } \mu<\mu^*,
\end{equation}
where $\alpha$ is replaced by $2$ if $\alpha>2$. This equation will define for us the conditions for which a phase transition is possible or impossible, for any value of $\alpha$, $\mu$, and $d$, by using the conditions for $\sigma$ in the 1DL model \cite{gori2017one}. If $\alpha\geq d$ ($\sigma\leq 0$, the green area in Fig. \ref{figure9} below the dotted $\mu=-1$ line) a GC is guaranteed to exist for arbitrarily small value of $\rho$. In other words, $\rho_c=0$ and there are no phase transitions. For $d=1$ and $d=2$ this region coincides with a return probability of $1$ \cite{ross1996stochastic}. 

For $d/2<\alpha<d$ ($0<\sigma<1$, the grey zone in Fig. \ref{figure9} below the dotted $\mu=-1$ line) a phase transition is possible, i.e., there is a possibility of a critical value $\rho_c$ which is between 0 and 1. A phase transition will be apparent in this case as $\rho$ is changed from $\rho<\rho_c$ to $\rho>\rho_c$. Finding an exact expression for the value of $\rho_c$ as a function of $\alpha$, $\mu$, and $d$ is an extremely difficult task which we will not attempt here. We note that a phase transition is possible in this region, but not guaranteed. For some values of $\alpha$ and $\mu$ we find $\rho_c=0$ whilst for others $\rho_c>1$. Fig. \ref{figure9} shows an example where phase transition was detected. 
For $\alpha\leq d/2$ ($\sigma\geq 1$, the red zone in Fig. \ref{figure9} below the dotted $\mu=-1$ line), a GC is guaranteed not to exists, i.e., $\rho_c>1$ \footnote{To be more precise, for a GC to exist in this region each pair of consecutive steps much be connected with an edge (corresponding to $C=1$ in the 1D lattice case). This will only be true if $f(\vb{x})$ is bounded such that $f(\lvert \vb{x}\rvert>L) = 0$ (meaning $\alpha=\infty$) and $\rho=1$.}.  

To summarize, in the weak LR region, where \mbox{$\sigma+1=d/\alpha$} (\ref{weaklr}),
\begin{equation}\label{musigma}
\mu<\mu^*
\begin{cases}
\begin{alignedat}{2}
     &\rho_c=0 \text{ (GC exists)}, \quad && \text{for } \alpha\geq d \\
     &0<\rho_c<1 \text{ possible}, \quad && \text{for } d/2<\alpha< d \\
     &\rho_c>1 \text{ (No GC)}, \quad && \text{for } \alpha\leq d/2.
\end{alignedat}
\end{cases}\\
\end{equation}

In the strong LR region (Fig. \ref{figure9}, above the dotted $\mu=-1$ line), the correspondence with the 1D lattice case is summarized by
\begin{equation}
    1+\sigma = \frac{1-\mu}{\alpha}, \quad \text{for } \mu>\mu^*.
\end{equation}
In complete analogy to the weak LR case, we find that
\begin{equation}\label{musigma2}
\mu>\mu^*
\begin{cases}
\begin{alignedat}{2}
     &\rho_c=0 \text{ (GC exists)}, \quad && \text{for } \alpha\geq 1-\mu \\
     &0<\rho_c<1 \text{ possible}, \quad && \text{for } \frac{1-\mu}{2}<\alpha< 1-\mu \\
     &\rho_c>1 \text{ (No GC)}, \quad && \text{for } \alpha\leq \frac{1-\mu}{2}.
\end{alignedat}
\end{cases}
\end{equation}
In Fig. \ref{figure9} these regions are separated by the black diagonal lines \footnote{Furthermore, without proving here, we note that the case of $\mu=\mu^*$ is characterized by interactions of the form $\frac{\log(\lvert i-j\rvert)}{\lvert i-j\rvert^\beta}$, for some power $\beta$.}. 

Along with the theoretical phase diagram, Fig. \ref{figure9} also presents a simulated construction of the phase diagram. There, we detect the existence of phase transitions without explicitly calculating the value of $\rho_c$ (a numerically expensive task), by finding the general region where the Second Largest Component (SLC) is maximal \cite{margolina1982size}. We calculate the SLC for only three values of $\rho$, 0.05, 0.5, and 0.95. If the maximal SLC is $<0.05$ we colored the corresponding squares green. If the maximal SLC is $>0.95$, we colored the corresponding squares red. If the maximal SLC is found in $\rho=0.5$, we colored the corresponding squared gold. These are the regions where phase transitions were detected.

An accurate way to estimate $\rho_c$ explicitly is to use the measure $Q_G$ proposed in \cite{gori2017one} 
\begin{equation}\label{QG}
    Q_G = \left\langle \frac{\sum_\mathcal{C} \mathcal{N}_\mathcal{C}^4}{(\sum_\mathcal{C} \mathcal{N}_\mathcal{C}^2)^2}\right\rangle,
\end{equation}
where $\mathcal{N}_\mathcal{C}$ is the number of nodes in cluster $\mathcal{C}$ and the averaging is done over the different realizations. $Q_G$ measures the spread of the cluster size, and acts as a Binder cumulant of the system \cite{binder1981finite}. When a GC is present, $Q_G\rightarrow 1$, otherwise $Q_G\rightarrow 0$. 
The curves of $Q_G$ as a function of $\rho$ for a different number of nodes, $N$, will cross near the critical $\rho_c$ (if one exists). Examples of such plots for various choices of $\alpha$ and $\mu$ are presented in the SM. Calculation of the critical $\rho_c$ using $Q_G$ for various values of $\alpha$ and $\mu$ is also presented in the SM. 

As we mentioned, in the SM we show the phase diagram of the system for $d=1$ to $d=9$. Note that for $d>3$, a GC can not emerge in the weak LR region, i.e., $\rho_c(d>4, \mu<\mu^*)>1$.

\textit{Discussion.}--In this work, we present a first and necessary step towards answering the following question: How do structures form in spatially scale-invariant networks with long-range interactions? This question is important as many real-world systems can display both of these properties. For example, bus stations in a country can have a scale-invariant distribution of distances, but some long-range bus lines between distance cities do often exist \cite{shanmukhappa2018spatial, wang2020analysing}. For a fully connected transport system, we need to understand how the proclivity of long-range lines, together with the scale-invariant distribution of distances between the stops, compete with each other to form a large structure. 

Our solution relies on the way we positioned the nodes in the metric space. Since we used a L\'evy flight procedure, which is a Markovian process that places the nodes sequentially, we were able to calculate the probability that any two steps are connected, and compare it to the known model of 1D percolation. This in turn produced a complicated and rich behavior of the system. First, we were able to find conditions for which a GC will always exist or will never exist. Then, we found conditions for which a phase transition is possible, and we detected these phase transitions using simulations. This combines techniques and behaviors from both long-range interaction theory and random walk theory in beautiful harmony. It is important to note, however, that we have no particular reason to suspect that real-world systems are produced by a similar L\`evy flight procedure, and a more complete theory that can handle heterogeneous scale-invariant placement of nodes that is not procedural or Markovian is needed.

Other than the conditions for which a GC emerges, we also found a transition from weak to strong long-range interactions. This is in line with a known behavior of Ising model with long-range interactions. A critical value $\sigma^*$ exists that separates the system into two cases, one where the critical behavior is the same as the short-range analog, and one where the LR interactions are relevant and the system has a phase transition with a peculiar LR critical behavior. In this model a third region, one where mean field universal behavior is expected, also exists \cite{fisher1972critical, dutta2001phase, defenu2017criticality}. This corresponds to our green regions in the phase-diagram plots and so it is expected that mean-field theory will be applicable there. 
Note that a crucial difference between our model and the Ising model is that the parameter $\alpha$ assumes the role of the parameter $d$ in the Ising case, making the phase-diagram plot continuous.

Many other properties of the network are crying out to be analyzed. For example, although we detect phase transitions in the predicated region of the phase diagram, we do not have a mathematical theory to explain how $\rho_c$ changes in this region, or why parts of this region do not exhibit a phase transition. We plan to study in the future other basic properties of this network, such as connectivity, clustering coefficients, centrality, and betweenness measures, and to analyze the critical exponents of these phase transitions. 



We thank Shlomo Havlin, Eli Barkai, and Stas Burov for fruitful discussions and helpful suggestions. This work was supported by the United States-Israel Binational Science Foundation (BSF), Jerusalem, Israel (Grant No. 2020255).

\bibliography{main_PRE}

\clearpage




\clearpage







\end{document}